\documentclass[12pt]{article}

\usepackage{verbatim} 
\usepackage{natbib} 
\usepackage{amsmath} 
\usepackage{amsbsy}
\usepackage{amssymb}
\usepackage{mathrsfs} 
\usepackage{lscape} 
\usepackage{graphicx}
\usepackage{epstopdf}
\usepackage{deluxetable}
\usepackage{ctable} 
\usepackage{fixltx2e} 

\newcommand{\beq}{\begin{equation}}
\newcommand{\eeq}{\end{equation}}
\newcommand{\beqa}{\begin{eqnarray}}
\newcommand{\eeqa}{\end{eqnarray}}

\newcommand{\aj}{{\it Astron. J.  }}
\newcommand{\apj}{{\it Astrophys. J. }}
\newcommand{\aap}{{\it Astron. Astrophys.}}
\newcommand{\mnras}{{\it Mon. Not.  R. Astron. Soc. }}
\newcommand{\apjl}{{\it Astrophys. J. Letters }} 
\newcommand{\apjs}{{\it Astrophys. J. Supplement }}
\newcommand{\nat}{{\it Nature }}

\def\araa{{\it Ann. Rev. Astron.  Astrophys. }}

\def\la{\lower.5ex\hbox{$\; \buildrel < \over \sim \;$}}
\def\ga{\lower.5ex\hbox{$\; \buildrel > \over \sim \;$}}

\begin{document}

\begin{center}
\noindent{\bf \large Recent progress in simulating galaxy formation from the largest to the smallest scales}

\bigskip
\noindent Claude-Andr\'e Faucher-Gigu\`ere
%\noindent Center for Interdisciplinary Exploration and Research in Astrophysics
%\noindent Department of Physics \& Astronomy
%\noindent Northwestern University, Evanston, IL , USA
\end{center}

%\bigskip

\noindent {\bf Abstract} 

\noindent Galaxy formation simulations are an essential part of the modern toolkit of astrophysicists and cosmologists alike. Astrophysicists use the simulations to study the emergence of galaxy populations from the Big Bang, as well as problems including the formation of stars and supermassive black holes. For cosmologists, galaxy formation simulations are needed to understand how baryonic processes affect measurements of dark matter and dark energy. Owing to the extreme dynamic range of galaxy formation, advances are driven by novel approaches using simulations with different tradeoffs between volume and resolution. Large-volume but low-resolution simulations provide the best statistics, while higher resolution simulations of smaller cosmic volumes can be evolved with more self-consistent physics and reveal important emergent phenomena. I summarize recent progress in galaxy formation simulations, including major developments in the past five years, and highlight some key areas likely to drive further advances over the next decade.

 \bigskip

\noindent Cosmology now has a standard model, in which of most of the mass is dark matter, the acceleration of the universe is due to dark energy, and in which tiny density perturbations in the early universe were seeded by inflation. 
In this ``$\Lambda$ cold dark matter'' ($\Lambda$CDM) model, described by just six parameters, baryons make up just five percent of the present-day energy density \citep[][]{2016A&A...594A..13P}.

Although the physical nature of baryons is much better understood than dark matter and dark energy, how primordial fluctuations eventually evolved into the galaxies that we use to map the universe in visible light remains a challenging problem at the frontiers of modern astrophysics. There are a few reasons that make galaxy formation one of the most active areas of astrophysical research today. These can be loosely grouped into two categories: astrophysics and cosmology. 

Astrophysicists want to know how galaxies formed and how they evolved because the diverse astronomical phenomena involved are interesting in their own right. For example, astrophysicists seek to understand the origins of galaxy properties, such as their masses, sizes, and colors, and why correlations between different properties (so-called ``scaling relations'') are observed. Astrophysicists are also interested in how galaxies came to be because it provides the context for understanding other problems, such as how stars and black holes formed in galaxies.

For cosmologists, the details of how galaxies assembled may not be of prime interest. However, cosmologists must know enough about galaxy formation physics to understand how their measurements are affected by how baryons interact with the dark sector. 
Cosmologists have so far been able to get away with a relatively crude understanding of how galaxies formed, usually relying on simulations containing only dark matter \citep[][]{2010ApJ...715..104H}, but this is changing. 
Indeed, upcoming experiments aiming to measure the equation of state of dark energy to better than one percent, such as the Large Synoptic Survey Telescope (LSST), the Euclid mission, and the Wide Field Infrared Survey Telescope (WFIRST), will require modeling baryonic processes with much greater accuracy. 
In particular, exploiting the statistical power of the weak lensing signal will require modeling the non-linear matter power spectrum at the level of one percent or better for scales corresponding to comoving wavenumbers $0.1 \lesssim k \lesssim 10$ $h$ Mpc$^{-1}$, where previous simulations have shown that baryonic effects can range from $\approx 1$ to $>10\%$ \citep[][]{2011MNRAS.415.3649V}. 
Since baryonic processes can substantially change the profiles of individual dark matter halos, they are also proving critical to constraining the properties of dark matter via dynamical measurements of galaxies, a point that I return to below. 

 \bigskip

\noindent {\bf Challenges and recent successes of large-volume simulations} 

\noindent Given the standard cosmology, the recipe for simulating galaxy formation is in principle straightforward: start with the right mix of dark matter, dark energy, and baryons, then integrate all the relevant evolution equations. The problem is of course that this brute force approach is well out of reach of computational capabilities, and this will remain the case for decades to come. Alternatives include semi-analytic techniques, in which baryonic processes are approximated with analytic prescriptions ``painted on'' dark matter-only simulations \citep[][]{1991ApJ...379...52W, 2001MNRAS.320..504S}, and semi-empirical methods in which observed galaxy populations are mapped to simulated dark matter distributions \citep[][]{2013ApJ...770...57B, 2013MNRAS.428.3121M, 2013MNRAS.435.1313H}. 

In what follows, I focus on recent progress using cosmological hydrodynamic simulations. Such simulations follow the coupled dynamics of dark matter and baryons starting from $\Lambda$CDM initial conditions. Simulations with volume sufficient to capture representative portions of the universe cannot resolve the interstellar medium (ISM) of galaxies in significant detail and are far from resolving the formation of stars. Such simulations typically have mass resolution $\sim 10^{6}$ M$_{\odot}$ and force resolution $\sim$1 kpc. These simulations therefore rely critically on ``subgrid'' models to capture processes internal to galaxies. To a large extent, advances in galaxy formation modeling are currently driven by the design and application of better subgrid models for the variety of crucial processes that cannot be explicitly resolved in cosmological simulations, such as star formation and stellar feedback, and supermassive black hole growth and active galactic nucleus (AGN) feedback.

The subgrid processes and their effects on resolved scales could in principle be so complex that they could not be captured by a manageable set of subgrid models. The first generations of cosmological hydrodynamic simulations in fact failed in many respects to produce realistic galaxy populations. They produced galaxies that were too massive and too compact \citep[][]{1994MNRAS.267..401N}. 
From earlier analytic and semi-analytic work \citep[][]{1974MNRAS.169..229L, 1986ApJ...303...39D, 1991ApJ...379...52W}, it was known that stellar feedback was important to produce realistic galaxy populations. Supernovae (SNe), in particular, can drive galaxy-scale outflows that eject gas from galaxies before it has time to turn into stars. 

The first attempts to include stellar feedback in cosmological simulations revealed that it is highly non-trivial. For example, when SNe are modeled by adding thermal energy to surrounding gas (``thermal feedback''), the feedback is inefficient because the energy is rapidly radiated away. This is one form of the ``overcooling problem,'' and is due to the fact that at the relatively coarse resolution of cosmological simulations, the energy from individual SNe is generally not sufficient to heat the gas enough to avoid rapid cooling. 
This is because the low resolution makes it difficult to resolve a multiphase ISM, where cooling times would be longer in the unresolved hot and tenuous gas phase. 
Several different methods have been developed to circumvent the overcooling problem. 
One method (``delayed cooling'') is to temporarily turn off radiative cooling to increase the efficiency of energy conversion into kinetic motion \citep[][]{2001ApJ...555L..17T, 2006MNRAS.373.1074S}. 
A variant (``stochastic heating'') keeps cooling on at all times, but temporarily stores feedback energy until a certain minimum heating temperature is reached \citep[][]{2012MNRAS.426..140D}. 
Another method (``hydrodynamically decoupled winds'') is to directly prescribe the desired velocity and mass loading of galactic winds \citep[][]{2003MNRAS.339..312S, 2006MNRAS.373.1265O}. 

Using such methods, several groups showed that stellar feedback models could be adjusted in ways that produce galaxy properties and demographics in much better agreement with observations, at least for galaxies of mass comparable to the Milky Way ($\sim L^{\star}$) or less \citep[][]{2010MNRAS.406.2325O, 2011ApJ...742...76G}, corresponding to dark matter halos of mass $M_{\rm h} \lesssim 10^{12}$ M$_{\odot}$. 
This confirmed that star formation-driven galactic winds could plausibly reconcile $\Lambda$CDM with observed galaxy populations. Simulations with galactic winds also enabled important advances in our understanding of how heavy elements synthesized in stars and stellar explosions were dispersed in the intergalactic medium.

These tentative successes stimulated much subsequent modeling of feedback in galaxy simulations, but it was recognized that the results were sensitive to model assumptions and thus that clear gaps remained in our understanding of how galaxies evolved. One influential simulation project, called OWLS (``OverWhelmingly Large Simulations''), demonstrated the dependence of simulation results on subgrid prescriptions particularly clearly by exploring more than fifty variations \citep[][]{2010MNRAS.402.1536S}. 

More recently, the trend has been to tune parameters of the subgrid prescriptions so that the simulations match certain basic observational constraints. The most basic constraint that all the simulations aim to reproduce is the galaxy stellar mass function, but additional properties such as galaxy sizes can break degeneracies between different models \citep[][]{2015MNRAS.450.1937C}. 
Two recent large projects, Illustris \citep[][]{2014MNRAS.444.1518V} and EAGLE \citep[][]{2015MNRAS.446..521S}, have followed this approach and produced simulated galaxy populations in boxes $\sim100$ Mpc on a side. In many respects, these recent simulations approximate observations well. In both simulations, stellar feedback is key to regulating star formation in galaxies below $L^{\star}$, but feedback from supermassive black holes must also be included to explain the properties of the most massive galaxies. 

The fact that different semi-analytic models \citep[][]{2003ApJ...599...38B, 2006MNRAS.365...11C} and cosmological simulations can explain galaxy stellar masses with the same basic ingredients is encouraging and suggests that feedback from stars and black holes are common elements of successful models, a point highlighted in a recent review of galaxy formation models \citep[][]{2015ARA&A..53...51S}. 
On the other hand, the fact that different variants of how the feedback is modeled produce similar galaxy mass distributions tells us that we have not yet converged on a unique theory of galaxy formation.

Fortunately, there are ways of distinguishing between different models. 
Once tuned to match basic observed properties, the simulations can be tested by comparing them with observations which were not used in the tuning. 
I highlight two sets of observations of particular significance for galaxy evolution. 

The first is the color distribution of galaxies. Galaxies are observed to have a bimodal color distribution, the ``blue cloud'' and the ``red sequence.'' 
The Illustris and EAGLE simulations were not tuned to match the color distribution of galaxies, so comparing with the observed color distribution is an important test. The original Illustris simulation predicted increasingly red colors with increasing galaxy mass, in qualitative agreement with observations, but rather large quantitative differences relative to observations from the Sloan Digital Sky Survey \citep[][]{2014MNRAS.444.1518V}. 
A more recent version of the Illustris simulation, IllustrisTNG, was designed in part to produce a better match to the observed galaxy color distribution \citep[][]{2018MNRAS.475..624N}. 
The EAGLE galaxies match the observed color distribution as a function of stellar mass about as well as IllustrisTNG \citep[][]{2017MNRAS.470..771T}, but both simulations appear to produce slightly too much residual star formation in some of the most massive galaxies and thus underpredict the observed tail of red galaxies.

The second is predictions for the gaseous halos of galaxies, known as the circum-galactic medium (CGM). Observations of the CGM (typically using quasar absorption lines, but also increasingly in emission) are powerful discriminants of galaxy formation theories because they directly probe the inflows and outflows that regulate galaxy growth. Comparing simulations with CGM observables has been an active area of research in the last few years, stimulated by the availability of rich data sets at both low and high redshifts. 
So far, the results of comparisons with observations have been mixed: they reveal both agreements and disagreements from which we are learning the limitations of the current models \citep[][]{2015MNRAS.449..987F, 2016MNRAS.462.2440T, 2017MNRAS.465.2966S}. 
Because the CGM provides a large number of different observational constraints (including absorption strengths and kinematics in different ions), it will continue to be a very fruitful approach to test galaxy formation models.

Even though they generally do not agree perfectly with observations and have their limitations, cosmological simulations provide extremely rich data sets that can be mined to provide insights into a wide array of science questions, ranging from the origins of galaxy morphologies to the chemical evolution of galaxies to the effects of galaxy evolution on the cosmic distribution of dark matter \citep[][]{2015MNRAS.454.1886S, 2016MNRAS.456.1235S, 2018MNRAS.475..676S}. 

\bigskip

\noindent {\bf Bridging cosmological and sub-galactic scales}

\noindent  In parallel to the developments summarized above, another line of research in galaxy formation modeling has gained momentum in the past few years. Until recently, detailed studies of star formation were largely decoupled from cosmological models because of the large separation of physical scales. Using a ``zoom-in'' approach, in which the large-scale cosmological environment is included at low resolution but in which the resolution is highly refined around galaxies of special interest, it is becoming possible to resolve scales comparable to individual star-forming regions. 
As a result, it is possible develop finer-scale subgrid models for galaxy formation simulations that are more directly tied to our understanding how stellar feedback acts on small scales. By tying subgrid models to constraints on the small-scale physics, we can break degeneracies between theories that agree on larger scales.

A compromise of highly refined simulations is that they cannot match the galaxy statistics provided by larger volume but lower resolution simulations. Nevertheless, several factors have motivated researchers to pursue cosmological zoom-in simulations and other types of highly resolved  models. 
First, some problems simply require higher resolution. These include resolving low-mass dwarf galaxies and the detailed internal structure of more massive galaxies. As dark matter-dominated systems, dwarf galaxies are important laboratories for constraining the properties of dark matter using astronomical observations. 
New observational facilities such as the Atacama Large Millimeter Array (ALMA), the James Webb Space Telescope (JWST) to be launched next year, and increasingly sophisticated integral field spectrographs on ground-based telescopes are mapping interstellar gas and stellar populations at high resolution in both large and small galaxies. 
Making full use of these observational capabilities requires simulations that resolve as much of the dynamical, thermodynamic, and chemical processes operating in galaxies as possible. 

Second, developing more explicit subgrid models for zoom-in simulations has stimulated fruitful cross fertilization between the fields of galaxy formation and star formation. Galaxy formation modelers are starting to draw more directly on the vast body of work on the physics of star-forming regions in constructing subgrid models \citep[][]{2012ApJ...759L..27P, 2012ApJ...745...69K, 2016ApJ...829..130R}. At the same time, researchers working on star formation physics can use galaxy formation simulations to include more realistic boundary conditions in their models \citep[][]{2016MNRAS.460.2297R}. The strengthening of ties between these two subfields of astrophysics has already enabled rapid progress, including some important advances that will be summarized below.

Third, researchers hope that the results of zoom-in simulations can be coarse grained to develop better subgrid models for large volume containing thousands of galaxies \citep[][]{2016MNRAS.462.3265D}. Such large-volume simulations will remain necessary to compute several important quantities of interest to both astrophysicists and cosmologists, including galaxy clustering, gravitational lensing by cosmic structures, microwave background anisotropies, and the Ly$\alpha$ forest. Large-volume simulations are also the best tool to capture the full range of galaxy evolution pathways. By studying the emergent outcomes of better resolved galaxy models anchored to higher resolution feedback models, simulators aim to reduce the number of parameters that must be tuned to reproduce observed galaxy populations. 

A priori, it is not clear that zoom-in simulations have sufficient resolution to meaningfully increase the predictive power of galaxy formation models. For example, state-of-the-art zoom-in simulations of Milky Way-mass galaxies have baryonic resolution elements of mass $\sim 10^{4}-10^{5}$ M$_{\odot}$ and spatial resolution $\sim10-100$ pc \citep[][]{2013MNRAS.428..129S, 2015ApJ...804...18A, 2016ApJ...827L..23W}, with on-going efforts aiming to improve these resolution parameters by one order of magnitude. 
By contrast, resolving the formation of individual stars would require a mass resolution better than 1 M$_{\odot}$. Moreover, the turbulent ISM has structure on scales orders of magnitude smaller than will be resolvable for the foreseeable future. 

The significance of the latest generation of cosmological zoom-in simulations is that they are starting to resolve a few key characteristic scales critical for capturing how stellar feedback operates in galaxies. 
In particular, zoom-in simulations of dwarf galaxies are now routinely evolved with resolution elements of mass $\lesssim 500$ M$_{\odot}$ \citep[][]{2017MNRAS.471.3547F} and are thus are often able to resolve the cooling radius of individual SN remnants (SNRs) in the ISM, corresponding to a swept up mass $M_{\rm cool} \approx 1,000$ M$_{\odot}$ (weakly dependent on ambient medium density and metallicity) \citep[][]{2015MNRAS.450..504M}. 
This is also sufficient to resolve the ISM into different star-forming regions. 
Together, these factors allow the simulations to much more accurately predict how SNe deposit energy and momentum in the ISM. Simulations of SN feedback have demonstrated that the clustering of SNe, inherited from the clustering of star-forming regions, is important to correctly model how different SNRs overlap and merge into large bubbles of hot gas \citep[][]{2016MNRAS.456.3432G, 2017MNRAS.470L..39F}. 
These hot bubbles can vent out of galaxies and appear important to generate galaxy-scale outflows carrying enough mass and energy to explain observed galactic winds.

Even in today's state-of-the-art zoom-in simulations, individual SNRs are typically not well resolved in higher-mass galaxies, but simulators have begun to adopt new solutions to the overcooling problem anchored to well-resolved SNR models. One solution, independently proposed by several groups \citep[][]{2014MNRAS.445..581H, 2015MNRAS.450..504M, 2015ApJ...802...99K, 2015MNRAS.451.2900K}, is to inject at the resolution scale of cosmological zoom-ins both the thermal energy \emph{and} radial momentum that each SNR would have had on that scale if its evolution had been resolved in the simulation. Injecting momentum in addition to thermal energy is important because the momentum of an SNR is boosted by an order of magnitude during the Sedov-Taylor phase. 
In practice, this is done by calibrating the momentum and residual thermal energy to the results of higher resolution SNR simulations. 
Another approach is to bypass modeling individual SNRs and instead use a subgrid model calibrated to match the properties of superbubbles produced by clustered SNe \citep[][]{2014MNRAS.442.3013K}. 

Many recent galaxy formation simulations have also begun to incorporate approximations for stellar feedback processes other than SNe, including radiation, stellar winds, and cosmic rays \citep[][]{2013ApJ...770...25A, 2013MNRAS.434.3142A, 2014MNRAS.445..581H, 2014ApJ...797L..18S, 2017ApJ...836..204N}. Although usually energetically subdominant relative to SNe, these other processes can be important because they couple differently to the ambient medium, they have different time dependencies, and they can interact non-linearly with one another. 
Simulations of massive galaxies are furthermore beginning to incorporate models of supermassive black hole growth and feedback that are increasingly anchored to high-resolution models of the small-scale physics \citep[][]{2017MNRAS.472L.109A, 2017MNRAS.470.4530W}.

\bigskip

\begin{figure}[!h]
\begin{center}
\includegraphics[width=0.6\textwidth]{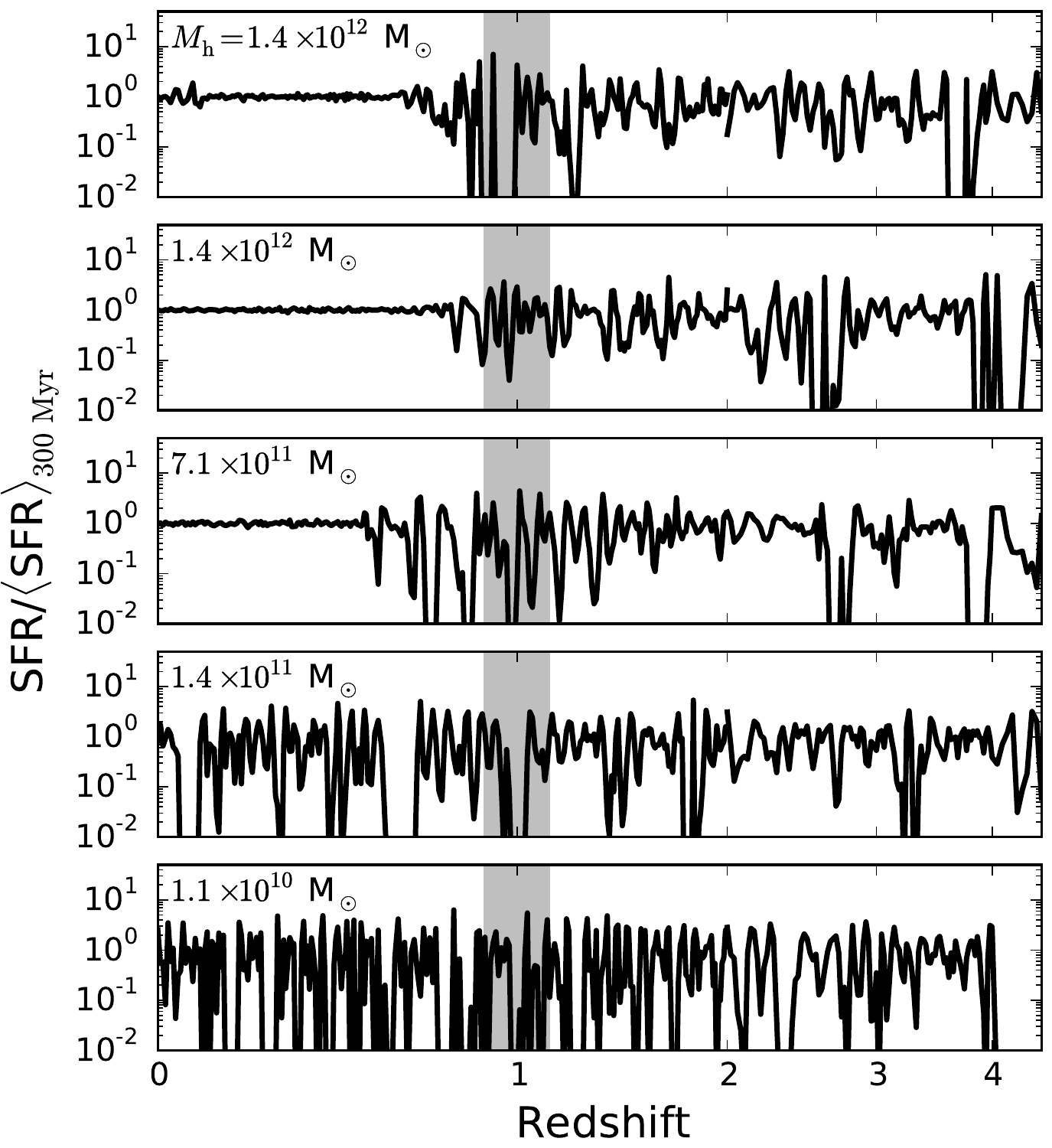}
\end{center}
\vspace{-0.2in}
\caption[]{\small 
{\bf Normalized SFR versus redshift for simulated galaxies from the FIRE project (running mean averaged over a period of 300 Myr).} The star formation histories are arranged in decreasing order of halo mass at redshift $z=0$, which is labeled at the top left of each panel. 
These simulations predict that all galaxies have bursty star formation histories at high redshift.  
The more massive galaxies settle into a time-steady mode of star formation at lower redshift but the dwarf galaxies continue to be bursty all the way to $z=0$. 
The approximate transition redshift $z\sim 1$ between bursty and time-steady star formation in massive galaxies is indicated by the grey bands. 
This transition in star formation variability corresponds to a gas morphology evolution from chaotic to a well-ordered disc configuration (see Fig. \ref{fig:gas_morphology}). 
The bursty star formation predicted by high-resolution simulations like these has important implications ranging from dark matter halos in dwarf galaxies to the growth of supermassive black holes. 
Adapted from \cite{2018MNRAS.473.3717F}. 
}
\label{fig:FIRE_SFHs_norm_NA} 
\end{figure}

\noindent {\bf Recent successes and predictions of zoom-in simulations}

\noindent Despite the approximations used, the latest generation of cosmological zoom-in simulations has produced promising results. One large zoom-in simulation campaign, in which I have played a role, is the FIRE project (for ``Feedback In Realistic Environments''). In the FIRE simulations, individual SNe are resolved in time and modeled by injecting both energy and momentum, as described above. The FIRE simulations also include approximations for photoionization, radiation pressure, and stellar winds, following the energetics and time dependencies from a standard stellar population synthesis model \citep[][]{1999ApJS..123....3L}. In these simulations, star formation is self-regulated by stellar feedback \citep[][]{2014MNRAS.445..581H} and galaxy-scale outflows emerge from the collective action of feedback processes acting on small scales \citep[][]{2015MNRAS.454.2691M}. 

Encouragingly, the FIRE simulations (and the more recent FIRE-2 variants using a new hydrodynamics solver) do a reasonable job of reproducing the observationally-inferred relationship between stellar mass and dark matter halo mass over more than seven orders of magnitude in stellar mass, up to $\sim L^\star$. 
In these simulations, a Kennicutt-Schmidt relation between the star formation rate (SFR) surface density and gas surface density ($\Sigma_{\rm SFR}$ vs. $\Sigma_{\rm g}$) roughly consistent with observations \citep[][]{1998ApJ...498..541K} also emerges from regulation by stellar feedback. 
These results from the FIRE simulations are significant because the subgrid models for stellar feedback were anchored to the physics of SNR evolution and the energetics for the feedback mechanisms were not adjusted to match observed galaxy masses. Moreover, the simulations did not switch off hydrodynamic interactions or gas cooling to increase the efficiency of feedback processes. 

Because there is large variance in how galaxies evolve, even at fixed final mass, the modest samples of galaxies simulated using the zoom-in technique (typically ranging from a single main galaxy to at most a few dozen halos) do not allow the kind of rigorous statistical comparisons with observed galaxy populations possible with large-volume simulations. 
Moreover, zoom-in simulations are generally evolved with resolution that increases with decreasing galaxy mass, since finer resolution elements can be afforded for lower-mass systems. 
As a result, numerical convergence has not yet been demonstrated uniformly across the full range of galaxy masses simulated with zoom-ins. 
There could therefore remain significant discrepancies with observations that would become clearer with larger and/or more uniform simulation samples. 
Rather than matching observations ``within the error,'' arguably the most important contribution of high-resolution simulations like those of the FIRE project is in making predictions for emergent behaviors unanticipated from large volume studies. 
In this respect, the FIRE simulations have produced some important predictions that were indeed unexpected, including by this author. 
Such predictions can be used to test the high-resolution simulations, and have stimulated new lines of research. 
\begin{figure}[!h]
\begin{center}
\includegraphics[width=0.99\textwidth]{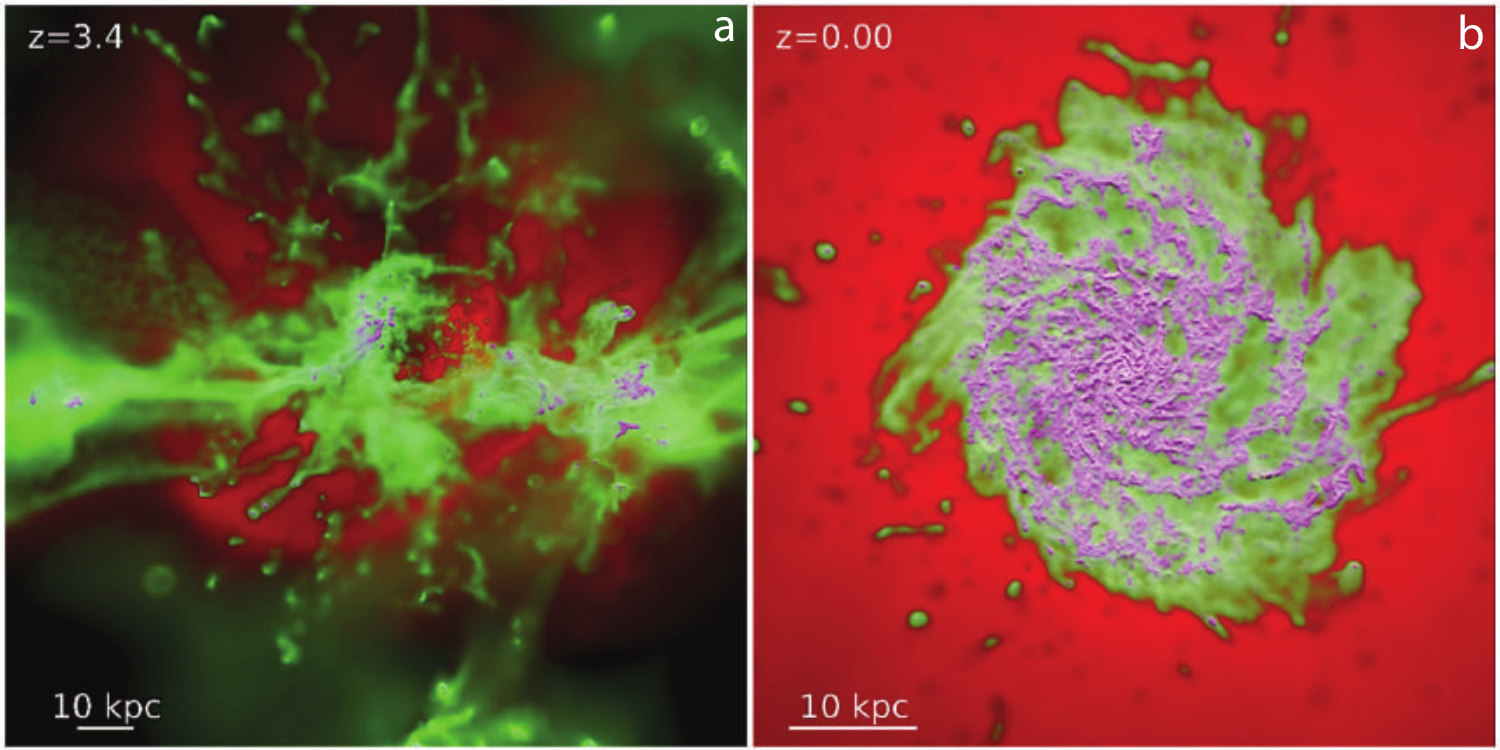}
\end{center}
\vspace{-0.2in}
\caption[]{\small {\bf Example gas distribution in a cosmological zoom-in simulation of a Milky Way-mass galaxy from the FIRE project}. Magenta shows cold molecular or atomic gas ($T \lesssim 1,000$ K), green shows warm ionized gas ($10^{4} \lesssim T \lesssim 10^{5}$ K), and red shows hot gas ($T\gtrsim10^{6}$ K). 
The gas is very clumpy and dynamic at high redshift (a; redshift $z=3.4$) and only later settles into a well-ordered rotating disc similar to spiral galaxies observed in the nearby universe (b; $z=0$). 
The transition in gas morphology occurs in tandem with the transition from bursty to time-steady star formation (Fig. \ref{fig:FIRE_SFHs_norm_NA}).  
} 
\label{fig:gas_morphology} 
\end{figure}

One key prediction of the FIRE simulations concerns the character of the star formation histories (SFHs) of galaxies. In most large-volume simulations to date, SFHs are relatively smooth in time and roughly determined by a competition between cosmological inflows and galactic winds \citep[][]{2012MNRAS.421...98D}. 
In contrast, the FIRE simulations predict SFHs that are much more time variable at high redshift, as well as in dwarf galaxies all the way to the present time (see Fig. \ref{fig:FIRE_SFHs_norm_NA}). 
Such bursty star formation is not unique to the FIRE simulations but appears generic in simulations that restrict star formation to high-density regions of the ISM \citep[][]{2012MNRAS.422.1231G, 2013MNRAS.429.3068T, 2015MNRAS.451..839D}. When the ISM is resolved into high-density clumps, star formation bursts can occur in rapid gravitational collapse events. The variability can be further enhanced by the explosive response of stellar feedback to local bursts of star formation. 
Figure \ref{fig:gas_morphology} shows how the emergence of well-ordered galactic discs correlates with the transition from bursty to time-steady star formation in massive galaxies. 
The predictions for SFR variability can be tested by measuring the SFRs of galaxies using light in bands that probe different timescales (e.g., recombination lines powered by young, massive stars vs. continuum emission including light from older stellar populations).

Several cosmological zoom-in simulations using different subgrid approximations have shown that bursty stellar feedback can transfer enough energy to the dark matter in the inner kiloparsec of dwarf galaxies to turn the cusps predicted by pure cold dark matter models into cored profiles \citep[][]{2010Natur.463..203G, 2013MNRAS.429.3068T, 2014MNRAS.437..415D, 2015MNRAS.454.2981C}. This effect is maximized in halos of mass $M_{\rm h} \sim 10^{10}-10^{11}$ M$_{\odot}$ as a result of a competition between the energy available from SN feedback and the depth of the gravitational potential. In contrast, galaxy formation simulations with smoother SFHs and standard cold dark matter do not produce such cored dark matter distributions \citep[][]{2014MNRAS.444.3684V, 2016MNRAS.457.1931S}. 
If observationally-inferred cores \citep[][]{2004MNRAS.351..903G, 2008AJ....136.2648D} are confirmed, e.g. by ruling out possible systematic effects in the modeling \citep[][]{2017arXiv170607478O}, knowing whether star formation is sufficiently bursty to explain the cores using baryonic effects will be critical to determine whether modifications to the standard cold dark matter paradigm are necessary. 

In more massive galaxies, bursty star formation has important implications for the growth of supermassive black holes and the emergence of galaxy-black hole scaling relations, such as the relation between black hole and stellar bulge masses ($M_{\rm BH}-M_{\rm bulge}$) \citep[][]{1998AJ....115.2285M}. 
For example, the FIRE simulations show that repeated gas ejection events by driven by bursty stellar feedback at early times can continuously deplete galactic nuclei of gas and delay the growth of central black holes relative to scaling relations observed in the local universe, and similar results have been found in other simulations as well \citep[][]{2017MNRAS.468.3935H, 2017MNRAS.472L.109A}. 

\bigskip

\noindent {\bf Conclusion and outlook}

\noindent Galaxy formation is far from solved, but the last five years have seen major advances in modeling using cosmological hydrodynamic simulations. These advances are enabling new insights into the variety of baryonic processes involved and their emergent outcomes. Large-volume hydrodynamic simulations are for the first time matching observed galaxy demographics at a level comparable to finely-tuned semi-analytic models, while higher resolution simulations are starting to resolve the ISM of individual galaxies and are making new testable predictions. 
Future progress will continue to be driven by both large volume and high resolution simulations. In fact, the synergy between the two approaches is likely to grow stronger as the high-resolution simulations are used to refine the subgrid models used in large volumes, and as these large volumes are exploited to investigate the implications for cosmology and large galaxy samples.
Before closing, I highlight some key areas where progress will be particularly fruitful going forward.

First, approaches to coarse grain the physics captured in high-resolution models into subgrid prescriptions remain somewhat {\it ad hoc} and it would be highly beneficial to develop more systematic methods. 
Second, the physical processes included in current simulations are incomplete and often rely on crude approximations. 
In this area, rapid progress is already underway using simulations that include combinations of magnetic fields, cosmic ray transport, radiation-hydrodynamics, and detailed chemistry networks, but the complexity of the problem guarantees that this line of investigation will remain open for the foreseeable future. 
Third, most simulation codes do not take full advantage of the supercomputing facilities available today. 
This is especially the case for highly zoomed in simulations, which often only scale well to a few hundred or thousand compute cores (out of hundreds of thousands cores on national supercomputers accessible to scientists). 
Moreover, the largest supercomputers increasingly rely on acceleration by graphics processing units (GPUs) or other many-core technology, but most current simulation codes are not yet designed to benefit substantially from these hardware accelerators. 
To some degree, progress in galaxy formation simulations has therefore been limited by the capabilities of the simulation codes and it will be important to improve both their scaling and hardware acceleration to make use of upcoming exascale facilities, which will become available in the next few years. 
This will be needed not only to simulate the astrophysics at significantly higher resolution and with a richer set of physical processes included self-consistently, but also to evolve hydrodynamic simulations with volume of multiple cubic gigaparsecs and trillions of resolution elements. 
Such simulations will be necessary to exploit the full information content of wide-field sky surveys of the next decade.

\bigskip

%\noindent {\bf Acknowledgments}

\noindent Correspondence and requests for materials should be addressed to the author. The author declares no competing financial interests.

\newpage
%\bibliography{references}

%\end{document}

\end{document}